\documentclass [12pt,elsart,epsfig]{article}     

\usepackage{epsfig}
\begin{document}
\begin{center}

{\Large \bf $I=0$, $C = -1$ mesons from 1940 to 2410 MeV}
\vskip 5mm

{A.V. Anisovich$^d$, C.A. Baker$^a$, C.J. Batty$^a$, D.V. Bugg$^c$, L. Montanet$^b$, 
 V.A. Nikonov$^d$, A.V. Sarantsev$^d$, V.V. Sarantsev$^d$, 
B.S.~Zou$^{c}$ \footnote{Now at IHEP, Beijing 100039, China} \\
{\normalsize $^a$ \it Rutherford Appleton Laboratory, Chilton, Didcot OX11 0QX,UK}\\
{\normalsize $^b$ \it CERN, CH-1211 Geneva, Switzerland}\\
{\normalsize $^c$ \it Queen Mary and Westfield College, London E1\,4NS, UK}\\
{\normalsize $^d$ \it PNPI, Gatchina, St. Petersburg district, 188350, Russia}\\
 }
\end {center}

\begin{abstract}
New Crystal Barrel data are reported for $\bar pp \to \omega \eta$
and $\bar pp \to \omega \pi ^0\pi ^0$ with $\omega$ decaying to
$\pi ^+\pi ^-\pi ^0$.
The $\omega \eta $ data confirm angular distributions
obtained earlier from data where $\omega \to \pi ^0 \gamma$.
The new $\omega \eta$ data provide accurate measurements of vector and
tensor polarisations of the $\omega$ and lead to considerable
improvements in masses and widths of $s$-channel
resonances.
A new $J^{PC} = 3^{+-}$ $I = 0$ resonance is observed with mass
$M = 2025 \pm 20$ MeV and width  $\Gamma = 145 \pm 30$ MeV.
Polarisation is close to zero everywhere
and tensor polarisations are large, as is the case also for
$\bar pp \to \omega \pi ^0$.

\end{abstract}

Earlier publications have reported studies of $\bar pp \to
\omega \eta$ [1] and $\bar pp \to \omega \pi ^0 \pi ^0$ [2]
for beam momenta 600 to 1940 MeV/c. That work concerned
all-neutral final states where $\omega$ decays to $\pi ^0 \gamma$.
Those data suffer from the disadvantage that much of the information
concerning $\omega $ polarisation is lost, because it is transferred
to the unmeasured polarisation of the decay photon.
In order to recover that information, we present here new data
where $\omega \to \pi ^+\pi ^- \pi ^0$.
In these data, the polarisation of the $\omega$ is determined
fully by the normal $\vec n$ to the decay plane of the $\omega$, as
explained in an accompanying paper on $\omega \pi ^0$ and $\omega
\eta \pi ^0$ final states [3].
Data for $\bar pp \to \omega \eta$ have also been reported by Peters [4].

Analyses of channels with quantum numbers $C = +1$, $I = 0$ and 1 [5-7]
have led to a spectrum of $s$-channel resonances consistent with
almost all of the expected $q\bar q$ states in the mass range 1900--2400
MeV.
The data presented here lead to major improvements in the identification
of resonances with $I = 0$, $C = -1$.
All the expected states are observed except for two $J^{PC} = 1^{--}$
states.
In that sector there is a problem separating $^3S_1$ and $^3D_1$
partial waves.

The present work follows closely the procedures described in
the accompanying paper on $I = 1$, $C = -1$, where data for final states
$\omega \pi ^0$, $\omega \eta \pi ^0$ and $\pi ^- \pi ^+$ are
discussed.
Most experimental details are common to that work,
so we refer to it for description of the experimental set-up
and many technical details.

The essential experimental problem is to minimise background from
$\eta \pi ^+\pi ^- \pi ^0$ in the $\omega \eta$ data and from
$\pi ^+\pi ^-\pi ^0\pi ^0\pi ^0$ in the $\omega \pi ^0 \pi ^0$ data.
An initial selection requires that both charged
particles are produced with centre of mass angle $\theta$,
$|\cos \theta | \le 0.65$, in order to avoid edge-effects in the
drift chamber.
Here $\theta$ is the lab angle of the $\omega$ with respect to the beam.
At least 11 digitisations are required in this chamber, with at
least one hit in the first three layers and at least one in the
last three.
In the preliminary selection, a 4C kinematic fit is required with
confidence level $>5\%$ for $\eta \pi ^+\pi ^- \pi ^0 $ or
$\pi ^+\pi ^-\pi ^0 \pi ^0  \pi ^0$.

After this initial selection, clear signals are observed for $\omega
\to \pi ^+\pi ^-\pi ^0$ in both sets of data, as shown in
Figs. 1(a) and (b);
all $\pi ^0$ combinations are included in the latter figure.
The background under the $\omega$ is fitted to a quadratic function of
mass and the $\omega$ is fitted to a Gaussian.
Events are selected for further processing in the mass range
760--804 MeV;
the few events where more than one $\pi ^+\pi ^-\pi ^0$ combination lies
in this interval are rejected. Surviving events are then subjected to a
kinematic fit to final states $\omega \eta$ or $\omega \pi ^0 \pi ^0$,
setting the $\omega$ mass to 781.95 MeV.

\begin{table} [htp]
\begin{center}
\caption {Numbers of selected events (including background)}
\begin{tabular} {ccc}
\hline
Beam momentum & $\omega \eta$ & $\omega \pi ^0 \pi ^0$ \\
(MeV/c)       &               &                         \\\hline
600           & 1139 & 5446   \\
900           & 4768 & 25071  \\
1200          & 2106 & 11510  \\
1525          & 612  & 4893   \\
1642          & 1209 & 11031  \\
1940          &  469 & 5241   \\\hline
\end{tabular}
\end{center}
\end{table}

The final selection of $\omega \eta $ and $\omega \pi ^0 \pi
^0$ requires a confidence level $(CL) > 10\%$ and greater than
that of any background channel.
As a minor refinement to check that the $\omega$ is well
reconstructed, it is required that $CL(\omega \eta ) > 0.8 \times
CL(\pi ^+\pi ^- \pi ^0 \eta)$ and $> 0.5 \times CL(\pi ^+\pi ^-
4\gamma )$. Corresponding cuts are applied in the selection of
$\omega \pi ^0 \pi ^0$ events. The
efficiency with which events pass the final kinematic fit is determined
as a function of $\pi ^+\pi ^-\pi ^0$ mass, and this is used to
evaluate the resulting background under the $\omega$. For $\omega \eta$
it increases steadily with beam momentum over the range 13 to 16\%; for
$\omega \pi ^0 \pi ^0$ it is 28--35\%. Here we also allow for the
contribution to background from `wrong' combinations of the spectator
$\pi ^0$ with $\pi ^+\pi ^-$. Numbers of selected events are shown in
Table 1.

\begin{figure}
\centerline{\hspace{0.2cm}\hskip 0.01cm\epsfig{file=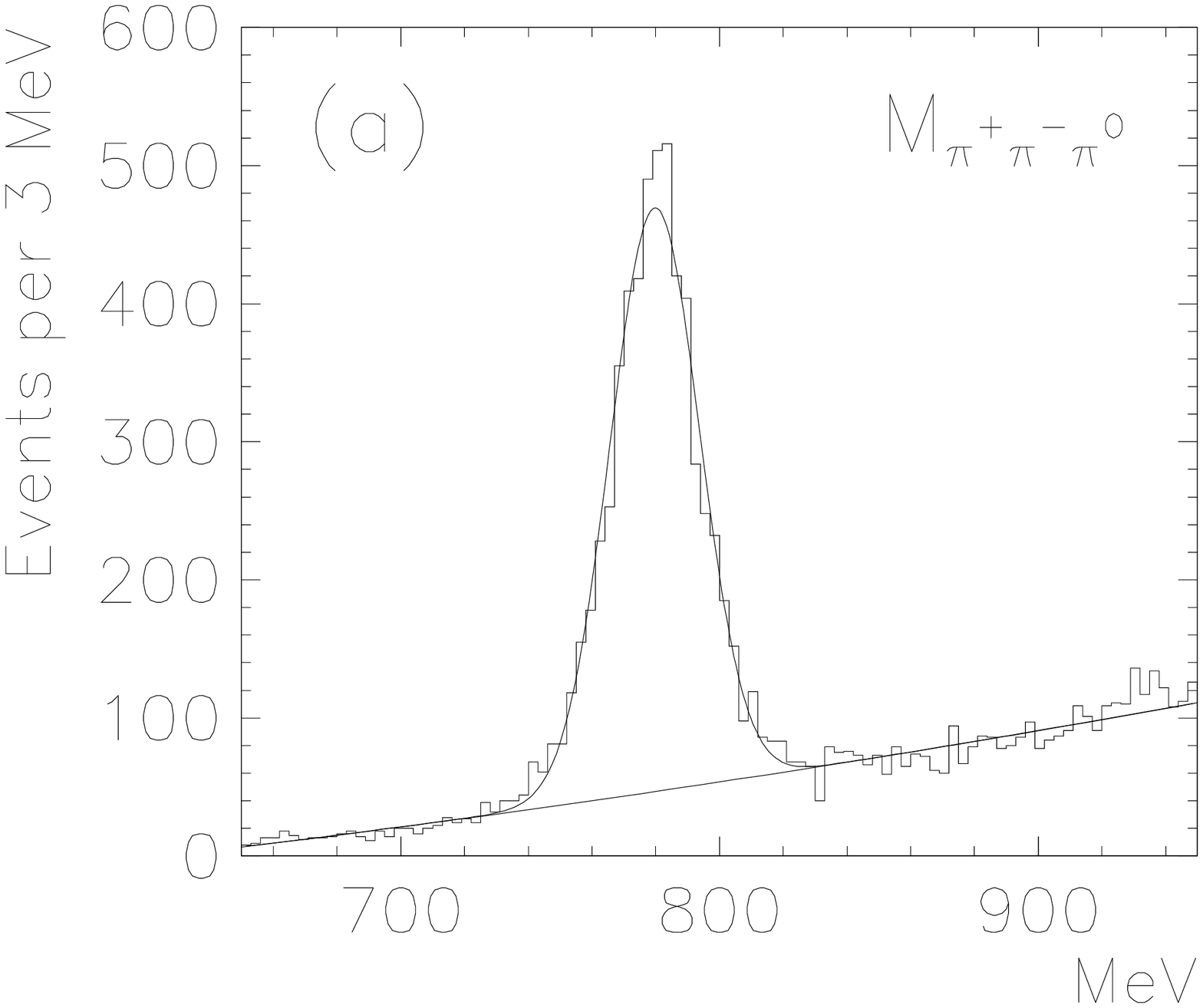,width=5.1cm}
            \hskip -0.01cm\epsfig{file=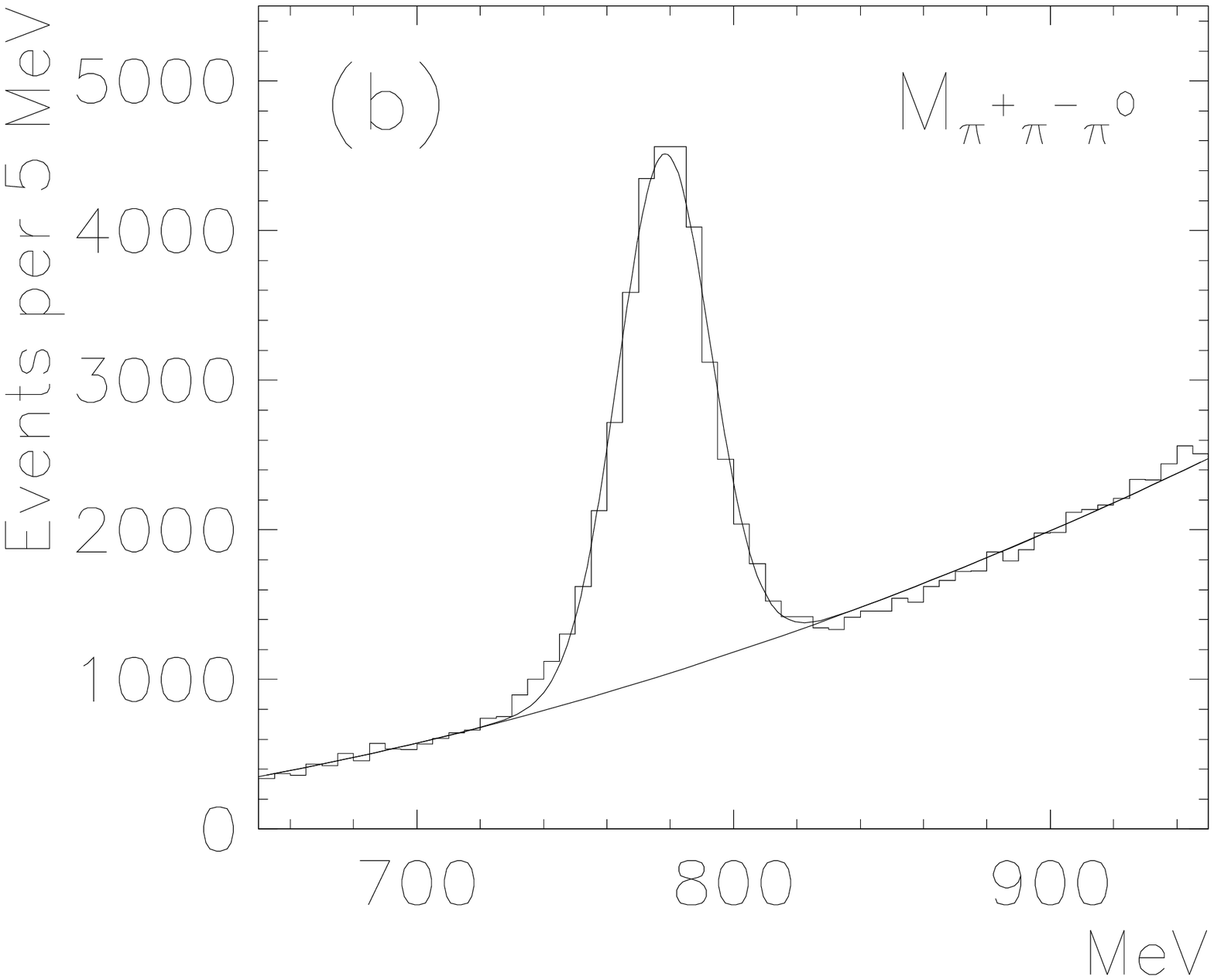,width=5.1cm}}
\vskip -5.14cm
\centerline{\hspace{0.2cm}\hskip 0.01cm\epsfig{file=F1A_WET.PS,width=5.1cm}
            \hskip -0.01cm\epsfig{file=F1B_WET.PS,width=5.1cm}}
\vspace{-1.1cm}
\centerline{\hspace{0.3cm}\hskip 0.01cm\epsfig{file=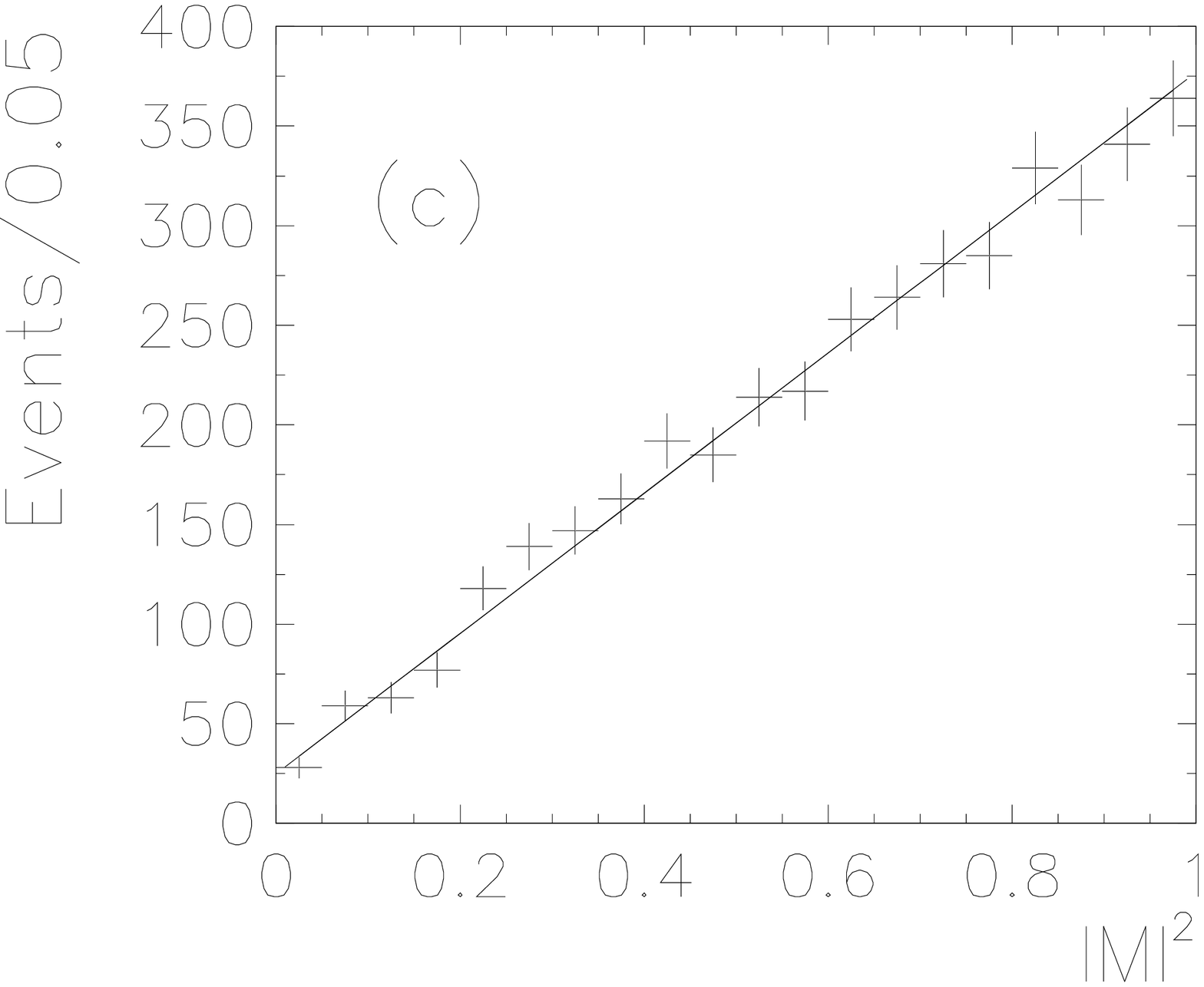,width=5.2cm}
            \hskip -0.01cm\epsfig{file=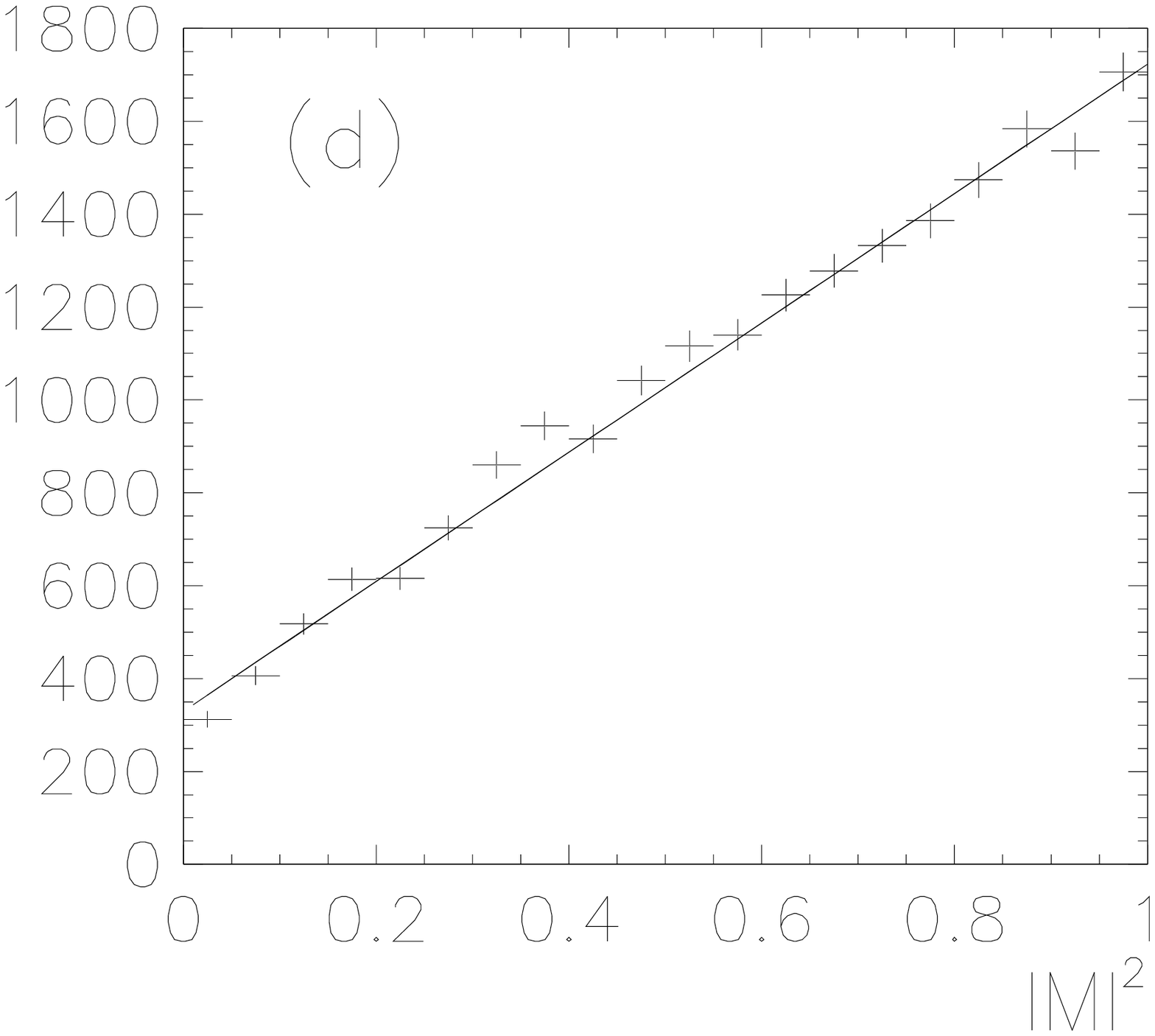,width=5.2cm}}
\vskip -5.24cm
\centerline{\hspace{0.3cm}\hskip 0.01cm\epsfig{file=F1C_WET.PS,width=5.2cm}
            \hskip -0.01cm\epsfig{file=F1D_WET.PS,width=5.2cm}}
\vspace{-1.5cm}
\centerline{\epsfig{file=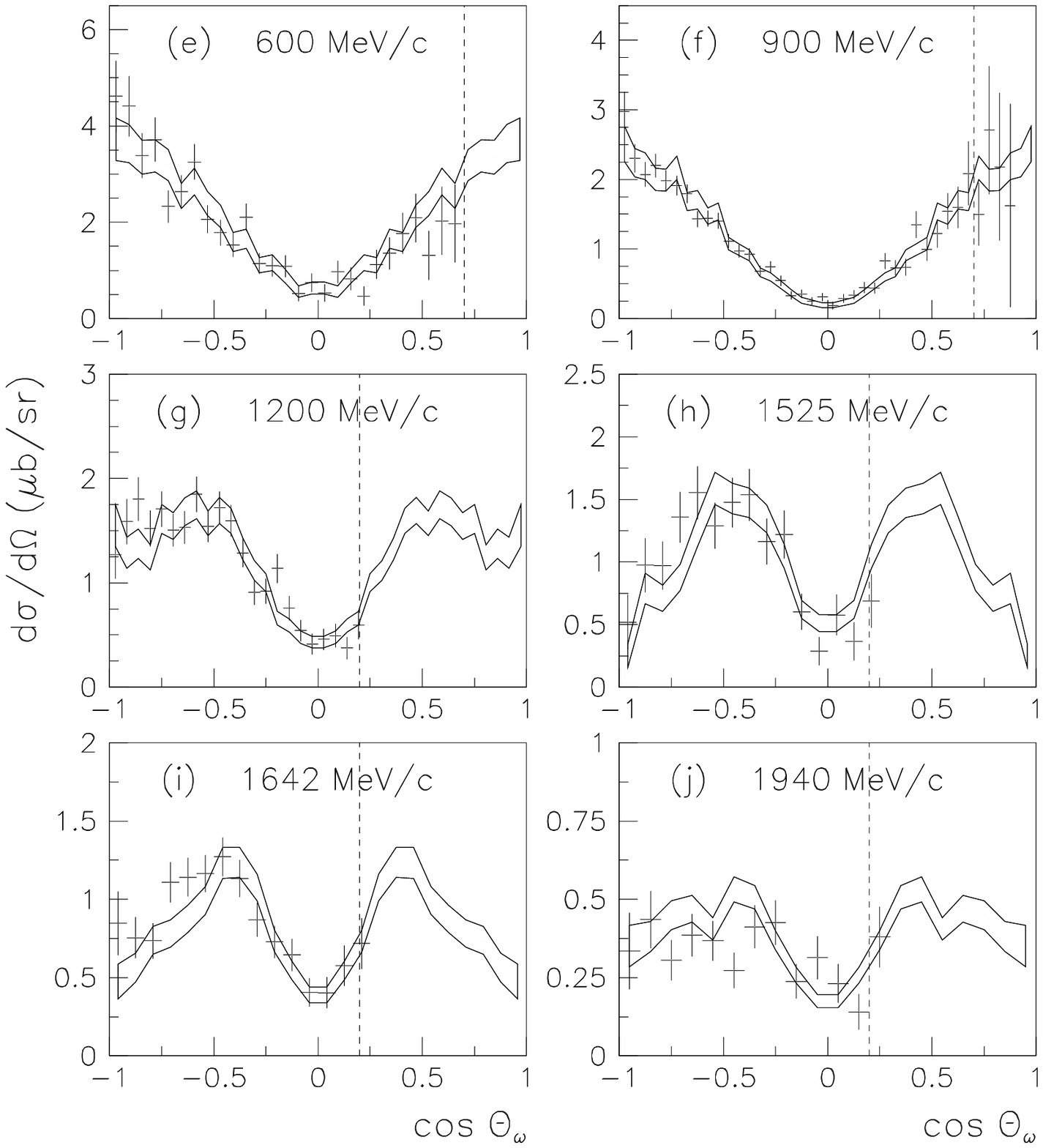,width=13cm}}
\vskip -130.25mm
\centerline{\epsfig{file=F1EJ_WET.PS,width=13cm}}
\vspace{-0.5cm}
\caption{$M(\pi ^+\pi ^- \pi ^0)$ from the preliminary data selection
for (a) $\omega \eta$ data, (b) $\omega \pi ^0 \pi ^0$ at 900 MeV/c;
the number of events v. the matrix element squared for
$\omega$ decay to (c) $\omega \eta $, (d) $\omega \pi ^0 \pi ^0$;
(e)-(j): curves show the error corridor for $\omega \eta$ differential
cross sections where $\omega$ decays to $\pi ^0 \gamma$;
corrections have been applied for angular acceptance; points with
errors show new results for $\omega$ decays to $\pi ^+\pi ^-\pi ^0$;
the vertical dashed lines show the cut-off in $\cos \theta
_{\omega }$ which has been used.}
\end{figure}

As explained in the accompanying paper [3],
the background may be estimated in a second way.
Decays of the $\omega$ are enhanced near
the edge of its Dalitz plot because of the momentum dependence of
the matrix element for $\omega$ decay.
Figs. 1(c) and (d) show plots of the number of events against the
square of this matrix element.
One sees straight lines with intercepts which provide another
estimate of the background; it agrees with the first within errors.
This background is included into the partial wave analysis described
below,
using Monte Carlo events which pass the data selection;
they are generated according to $\pi ^+\pi ^- \pi ^0 \eta $ or
$\pi ^+\pi ^-\pi ^0  \pi ^0 \pi ^0$ phase space.

We now compare angular distributions for present $\omega \eta$ data
with those from all-neutral data where $\omega \to \pi
^0\gamma$.
Figs. 1(e)-(j) show error corridors through the
latter data.
Points with errors are superposed from present data.
(Both sets of data are corrected for acceptance).
The absolute efficiency for
tracking charged particles has a significant uncertainty;
it is sensitive to
precise cuts on the number of layers and the $\chi ^2$
for the fit to a helix.
Therefore, the absolute scale for present charged data is normalised to
that for  all-neutral data.

Particles are thrown forwards in the lab system by the Lorentz boost
due to beam momentum.
In consequence, $\omega$ are detected efficiently only in the
backward hemisphere in the centre of mass.
This is adequate, since conservation of C-parity demands that the
production angular distribution is symmetric forwards and backwards
with respect to the beam.
Figs. 1(e)--(j) show that the shapes of angular distributions agree
well for the backward hemisphere between the two sets of data.
This agreement, seen particularly clearly at 900 MeV/c where statistics
are highest, is a valuable cross-check on experimental techniques.
In the forward hemisphere, we reject events where the acceptance for
the $\omega$ drops rapidly.
This requires a cut $\cos \theta _{\omega}< 0.7$ at 600 and 900 MeV/c,
$\cos \theta _{\omega}< 0.2$ at higher momenta.
These cuts reject only $\sim 10\%$ of selected events.

\begin{figure}
\vskip -6mm
\centerline{\epsfig{file=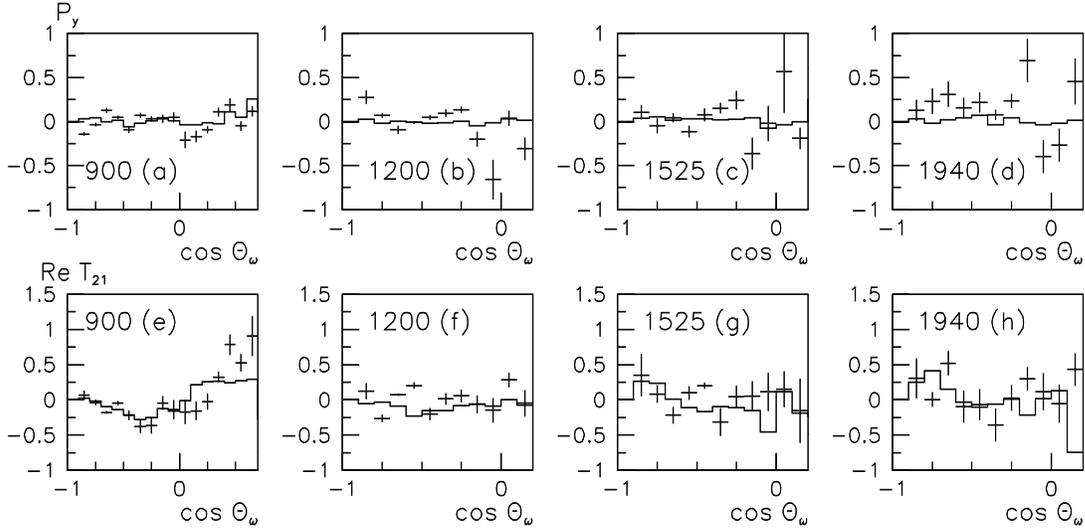,width=15cm}}
\vskip -150.25mm
\centerline{\epsfig{file=f21_wet.ps,width=15cm}}\vskip -78mm
\caption{ Vector polarisation $P_y$ at (a) 900, (b) 1200, (c) 1525 and
(d) 1940 MeV/c compared with the partial wave fit (histogram);
(e)--(h) $Re~T_{21}$ at the same momenta. }
\end{figure}

Vector and tensor polarisations of the $\omega$ are determined
following the procedures described in the accompanying paper [3].
We discuss first results for the two-body final state $\omega \eta$.
Fig. 2 shows values of vector polarisation $P_y$ and also $Re~T_{21}$
at four momenta;
Fig. 3 shows values of $Re~T_{22}$ and $T_{20}$.
Tensor polarisations are large.
Experimental values of $Im~T_{21}$ and $Im~T_{22}$ are consistent with
zero, as predicted theoretically.

\begin{figure}
\vskip -13mm
\centerline{\epsfig{file=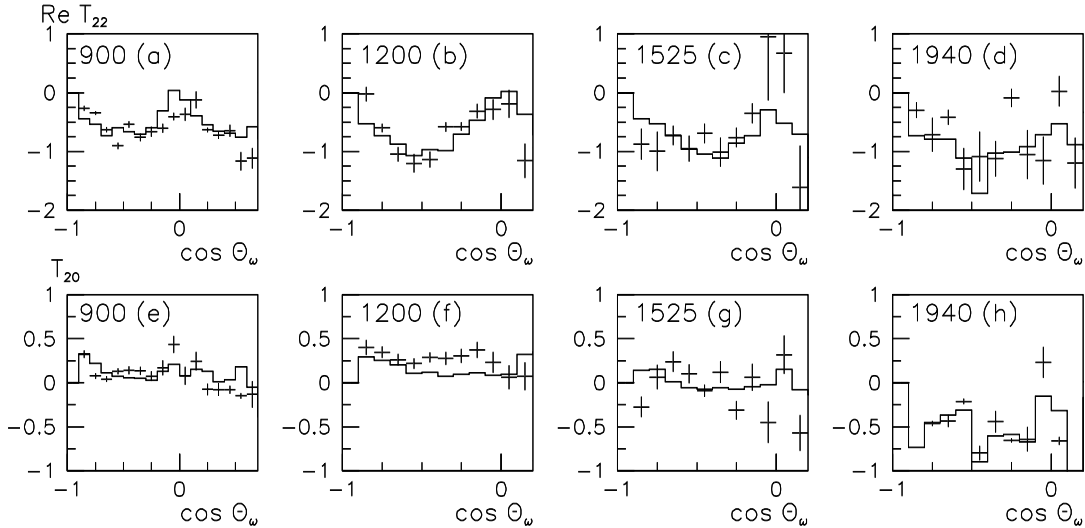,width=15cm}}
\vskip -150.25mm
\centerline{\epsfig{file=f31_wet.ps,width=15cm}}
\vskip -78mm
\caption{(a)--(d) $Re~T_{22}$ and (e)--(h) $T_{20}$ at 900, 1200,
1525 and 1940 MeV/c, compared with the partial wave fit (histograms).}
\end{figure}

We turn now to the partial wave analysis. This follows precisely
the lines described for $I = 1$, $C = -1$ [1--3].
In outline, partial wave amplitudes are described as a sum
\begin {equation}
f = \sum _i \frac {g_i\exp(i\phi _i)B_L(q)B_{\ell}(p)}
                  {M_i^2 - s - iM_i\Gamma _i }
\end{equation}
over $s$-channel resonances of constant width; fitted parameters
are masses and widths, coupling constants $g_i$ and phases
$\phi _i$.
Blatt-Weisskopf centrifugal barrier factors
$B_{\ell }$ and $B_L$ are included for production with
orbital angular momentum $\ell $ in the
$\bar pp$ channel and decay with orbital angular momentum $L$;
$p$ and $q$ are centre of mass momenta in $\bar pp$ and meson channels
respectively. We adopt a radius of 0.83 fm for the centrifugal barrier
radius in all partial waves up to angular momentum 3, as determined in
Ref. [5];
this radius is increased to 1.1 fm for higher partial waves.

The spectroscopic notation is described in detail in the
accompanying paper [3].
All triplet partial waves may couple in either
the $\bar pp$ entrance channel or in the decay to orbital angular
momentum $L = J \pm 1$. A resonance is expected to have the same phase
$\phi$ in these coupled channels, through rescattering between the two
channels. For present data there is an accurate determination of the
relative phase only for the two $2^{--}$ states; it is $-5.3 \pm
17.6^{\circ}$ for the lower one and $1.7 \pm 17.6^{\circ }$ for the
second. These are both consistent with zero, as expected. The same
result was found also in Refs. 3 and 5, with somewhat better errors.
Coupling constants for coupled channels are therefore fitted to a real
ratio $r = g_{L=J+1}/g_{L=J-1}$.

Table 2 shows parameters of fitted resonances.
The information from the polarisation of the $\omega$
leads to major improvements in the $\omega \eta$ channel compared
with the earlier work of Ref. [1].
For this channel, all resonances are now well determined in mass
and width, except for the $3^{--}$ state at 2285 MeV.
Columns 5 and 7 of Table 2 show changes in log likelihood when
each resonance is omitted from the fit and remaining resonances
are re-optimised. For convenience, columns 6 and 8 show
corresponding values from the earlier analyses. One sees
immediately a considerable increase in the significance of most
resonances.

\begin{table} [htp]
\begin{center}
\caption {Resonance parameters from a combined fit to $\omega \eta$ and
$\omega \pi ^0 \pi ^0$, using both
$\omega \to \pi ^0 \gamma$ and $\omega \to \pi ^+\pi ^- \pi ^0$ decays.
Values in parentheses are fixed.
Values of $r$ are ratios of coupling constants
$g_{J+1}/g_{J-1}$.
Columns 5 and 7 show changes in
$S = $log likelihood when each resonance is omitted from this fit and
others are re-optimised; columns 6 and 8 show a comparison with
previous work.} \vskip 2mm
\begin{tabular}{cccccccc}
\hline
$J^{PC}$ & Mass $M$ & Width
$\Gamma$ & $r$ & $\Delta S$ & Previous & $\Delta S$ & Previous \\
        & (MeV)  & (MeV) &  & $(\omega \eta )$ &
        $\Delta S (\omega \eta )$ & $(\omega \pi \pi )$ &
        $\Delta S (\omega \pi \pi )$ \\ \hline
$1^{+-}$ & $1965 \pm 45 $ & $345 \pm 75$  & -$0.20 \pm 0.17$ &
   1808 & 209  & 700 & 143\\
$1^{+-}$ & $2215 \pm 40$ & $325 \pm 55$ & $4.45 \pm 1.16$ &
    272 & 76 & 1020 & 360 \\
$3^{+-}$ & $2025 \pm 20$ & $145 \pm 30$ & $0.60 \pm 0.49$ &
    341 & - & 234 & - \\
$3^{+-}$ & $2275 \pm 25$ & $190 \pm 45$ & -$0.74 \pm 0.60$ &
    253 & 17 & 140 & 240 \\\hline
$1^{--}$ & $1960 \pm 25$ & $195 \pm 60$ & $2.4 \pm 0.45$  &
   1258 & 281 & 393 & 1515 \\
$1^{--}$ & $2205 \pm 30$ & $350 \pm 90$ & $-3.0 \pm 1.8$  &
    180  & 53 & 1410 & 1021 \\
$2^{--}$ & $1975 \pm 20$  & $175 \pm 25$ & $0.60 \pm 0.10$ &
    948 & 34 & 1203 & 478\\
$2^{--}$ & $2195 \pm 30$ & $225 \pm 40 $ & $0.28 \pm 0.59$ &
    160 & 94 & 356 & 590\\
$3^{--}$ & $1945 \pm 20$  & $115 \pm 22$ & $0.0 \pm 1.0$ &
    752 & 230 & 5793 & 595 \\
$3^{--}$ & $2285 \pm 60$ & $230 \pm 40$ & $1.4 \pm 1.0$ &
     98 & 46 & 1156 & 510 \\
$3^{--}$ & $2255 \pm 15$ & $175 \pm 30$ & $\sim 50$     &
    145 & 187 & 1436 & 695    \\
$4^{--}$ & $2250 \pm 30$ & $ 150 \pm 50$ & (0)          &
     70 & - &  -  & 35 \\
$5^{--}$ & $\sim 2250$ &$ 320 \pm 95$ & (0)  &
    677 &  - & 244 &  - \\\hline
\end{tabular}
\end{center}
\end{table}

Analysis of the $\omega \pi ^0 \pi ^0$ channel gives less precise
results for several reasons.
One is that the $\omega$ polarisation information is distributed
over 3-body phase space, though it is all used in the maximum
likelihood fit.
The decay channels which are significant are $\omega f_2(1270)$,
$\omega \sigma$ and $b_1(1235)\pi$, see Ref. [2]; there is also
a contribution from $f_0(1500)\omega$ at the highest two momenta.
The $b_1\pi$ channel is weak.
The $\omega \sigma$ channel is strong for several partial
waves, particuarly $J^{PC}=1^{--}$;
here $\sigma$ stands for the $\pi \pi$ S-wave amplitude, as
parametrised by Bugg, Sarantsev and Zou [8].
There is a problem concerning the treatment of the $\omega \sigma$
amplitude.
In $\pi \pi$ elastic scattering, there is an Adler zero
close to threshold, as a consequence of the nearly massless pion.
In coupling to heavy channels, $\bar pp$ and $\omega$, it is
not clear whether the Adler zero should be present in the
amplitude or not.
Some examples are known where the Adler zero is definitely absent,
e.g. $J/\Psi \to \omega \sigma$ [9].
We have explored both possibilities for every partial wave, choosing
the better alternative for every resonance.
This leads to considerable but unavoidable flexibility in the fit.

A related problem is that there are large interferences between
$\omega \sigma$ and $\omega f_2$. The precise phases of these
amplitudes affect masses of fitted $s$-channel resonances;
masses can shift with fitted phases in such a way as to
keep changes in log likelihood small. It leads to rather large
errors in resonance parameters for those partial waves where
$\omega \sigma$ contributions are big.
A final consideration is that background is fairly high for
$\omega \pi ^0 \pi ^0$ data.
Although the 3-body  data demand
large contributions to many partial waves, requiring
the presence of resonances, they lead to rather
imprecise determinations of resonance parameters in many
cases.

\begin{figure}
\vskip -20mm
\centerline{\epsfig{file=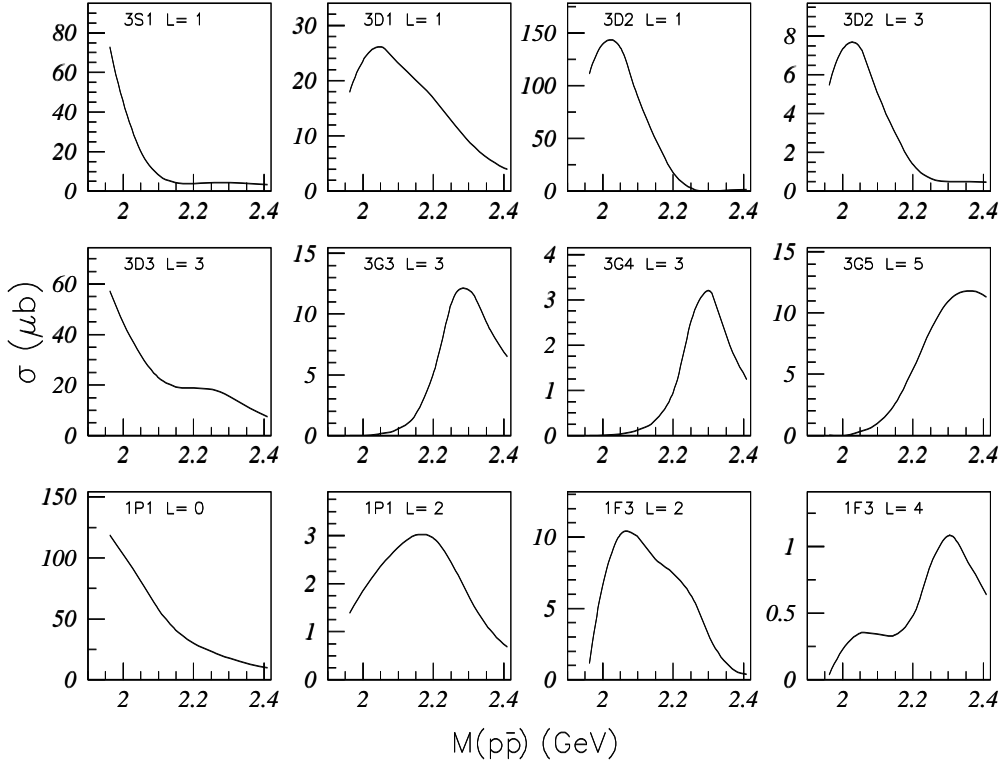,width=15cm}}
\vskip -120.35mm
\centerline{\epsfig{file=F4_WET.PS,width=15cm}}
\vskip -6mm
\caption{Intensities of partial waves fitted to $\omega \eta$ data;
$L$ are orbital angular momenta in the decay, and $3S1$, for example,
denotes $\bar pp ~^3S_1$.}
\end{figure}

We now discuss individual partial waves, beginning with singlet
states.
Their intensities in the $\omega \eta$ data as a function of mass are
shown in Fig. 4.
Intensities of the strong contributions to $\omega \pi \pi$ data are
shown in Fig. 5.
There are some changes compared with Figs. 5 and 6 of Ref. 2.
In several cases, these changes arise directly from better
determinations of $r$ parameters, by switching of amplitudes
between different $L$ values in the final state for a given
$J^P$.

\begin{figure}
\vskip -20mm
\centerline{\epsfig{file=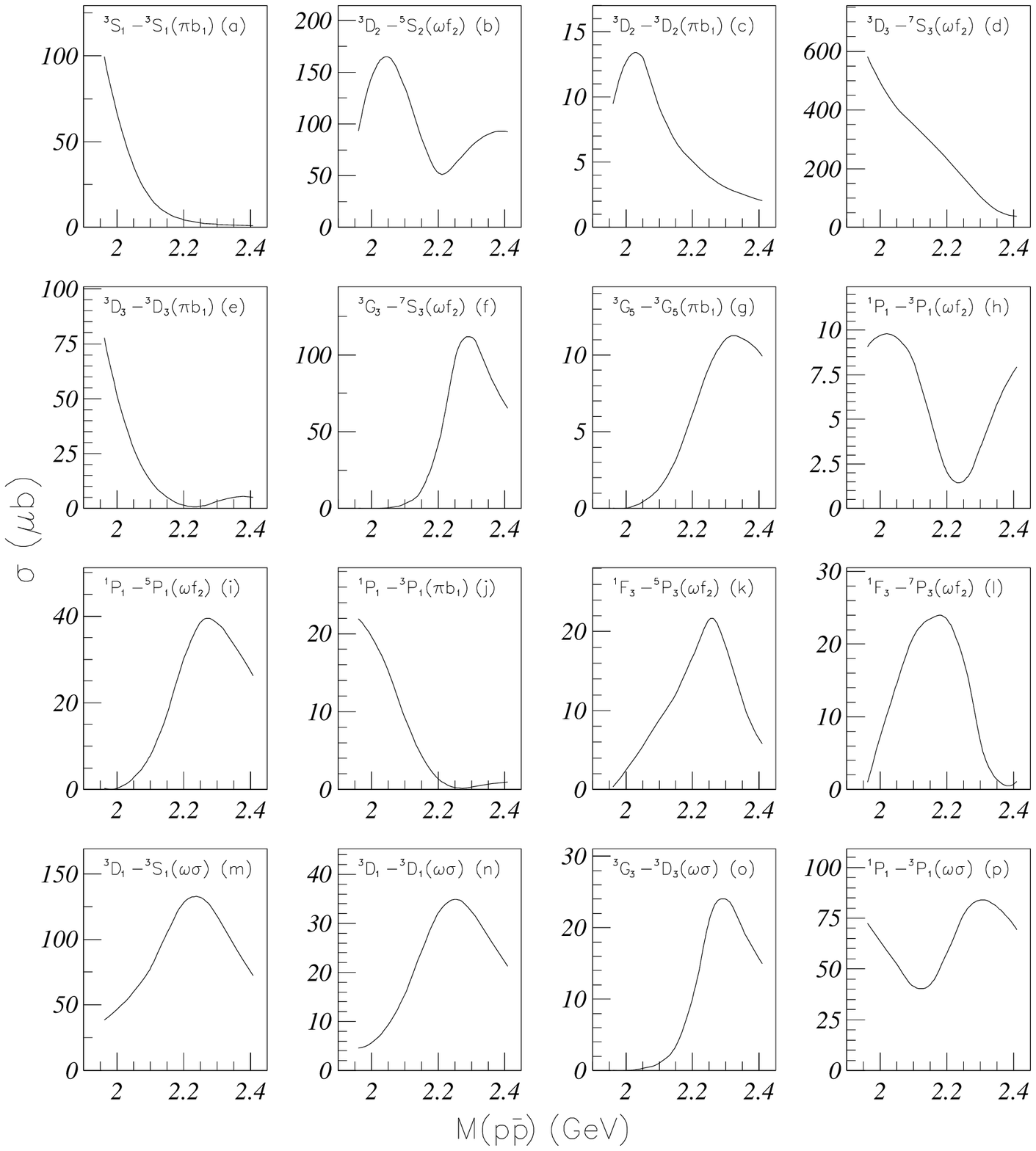,width=15cm}}
\vskip -165.35mm
\centerline{\epsfig{file=F5_WET.PS,width=15cm}}
\vskip -6mm
\caption{Intensities of the large partial waves fitted to $\omega \pi
^0\pi ^0$ data.
The spectroscopic notation is discussed in the text.}
\end{figure}

The spectroscopic notation of Fig. 5 is illustrated by that for panel
(b) as an example.
There, $^3D_2 \equiv ~ ^{2s + 1}\ell _J$ denotes the initial $\bar pp$
state, where $\ell = 2$ combines with spin $s = 1$ to make $J = 2$;
$P = (-1)^{\ell }$ and $C = (-1)^{\ell + s}$.
The $J = 2$ states may decay with $L = 0$, 2 or 4.
In almost all cases, only the lowest $L $ is significant for
$\omega \pi ^0 \pi ^0$ data, because of the centrifugal barrier.
In Fig. 5(b), $^5S_2 \equiv ~ ^{2s' + 1}L_J$ denotes the final
state $\omega f_2$ with $s' =  2$, $L =   0$, $J = 2$.
It is usually necessary to include all available $s'$ values for the
lowest $L$.
Examples are Figs. 5(h) and (i), where $s' = 1$ and 2 respectively,
combining with $L = 1$ to make $J = 1$; Clebsch-Gordon coefficients
tend to favour larger values of $s'$.

For $J^{PC} = 3^{+-}$, the present data reveal a new resonance
which escaped detection in the earlier analyses.
It lies at $2025 \pm 20$ MeV with $\Gamma = 145 \pm 30$ MeV.
A $q\bar q$ state is expected close to this mass as the singlet
partner to $^3F_2$, $^3F_3$ and $^3F_4$ states observed clearly
in Ref. [5].
Spin-spin splitting is believed to come from a $\delta$
function at zero radius [10]. In the absence of such splitting in
$F$-waves, the singlet state is expected to lie at the centroid
of the triplet states weighted by their multiplicities, namely
at $2024 \pm 5$ MeV; this agrees with the present result.

A second $3^{+-}$ state is observed at $2275 \pm 25$ MeV.
This compares with the prediction of $2292 \pm 9$ MeV
from triplet states.
Both of these $3^{+-}$ states are determined very largely by
the $\omega \eta$ data.
Although they make sizeable contributions to $\omega \pi ^0 \pi ^0$
data, errors are large there for masses and widths.

For $J^{PC} = 1^{+-}$, two states are observed at 1965 and 2215 MeV.
The accuracy of the mass determination of the lower state
comes mostly from $\omega \eta$,
although there is a large contribution in $\omega \pi ^0 \pi ^0$ too.
The error is sizeable, because this resonance is at the bottom
of the available mass range.
The upper state is required by the strong
$^5P_1(\omega f_2)$ contribution of Fig. 5(i). Both masses are slightly
lower than those from Refs. [1] and [2], as a consequence of the new
$\omega$ polarisation information.
The $q\bar q$ centrifugal barrier is
weaker for $1^{+}$ states than for $3^{+}$, so it is to be expected
that $1^+$ will resonate somewhat lower than $3^+$, as is observed.
The lower $1^{+-}$ state now appears in $\omega \eta$ almost purely
with  $L = 0$ decays; that is a large change from the earlier result
of Ref. [1].
The new polarisation data for the $\omega$ have provided a major
improvement in the determination of parameters $r = g_{L=J+1}/
g_{L=J-1}$, which describe the ratio of coupling constants $g$ for
$L = J \pm 1$.

We turn now to triplet partial waves, beginning with the $1^{--}$
sector, which is the most difficult.
For $J^{PC} = 1^{--}$, there are large
contributions in $\omega \pi ^0 \pi ^0$ and these data are mostly
responsible for fixing the upper resonance at 2205 MeV. The $\omega
\eta$ data do however produce the best determination of the lowest
$1^{--}$ resonance at 1960 MeV; it is weaker in $\omega \pi ^0 \pi ^0$,
though clearly visible in the $b_1(1235)\pi$ channel, Fig. 5(a). It
couples strongly to $^3D_1$, where its mass is best determined; it is
probably the radial excitation of $\omega (1650)$.

Four $1^{--}$ states are expected in this mass range: two $^3S_1$
and two $^3D_1$.
Unfortunately, $^3S_1$ and $^3D_1$ are not well
separated for remaining $1^{--}$ states in the absence of data from a
polarised target.
From a comparison with $I = 1$, $C = -1$ [3] where
data are available from a polarised target for the final state $\pi
^-\pi ^+$, higher resonances are expected around 2110--2150 and 2230
MeV and at a higher mass 2350--2400 MeV.
It is possible that the state
we report at 2205 MeV is a blurred combination of two states in the
mass range 2110-2230 MeV.
It has a large $\omega \sigma$ contribution,
shown in Figs. 5(m) and (n), and this contribution is somewhat
flexible, as discussed above.
This resonance does have quite a large
$^3D_1$ amplitude, and is consistent with the $^3D_1$ state expected
around 2230 MeV.
We have attempted to put two $1^{--}$ resonances into
the mass range 2100--2230 MeV, but they collapse to a single state.

In earlier work of Refs. [1] and [2], there was evidence for a further
$1^{--}$ state at $2300 \pm 45$ MeV.
Although an improvement in log likelihood of $\sim 300$ is possible
by including this state, there is no well defined optimum for its mass
and width. Therefore it is omitted from the final fit. Omitting it
has little effect on other partial waves.

For $J^P = 2^-$, two resonances are definitely required for two reasons.
Firstly, the phase variation visible in the Argand diagram of Fig. 7
below for $^3D_2~L=1$ is larger than can be provided by a single
resonance. Secondly, the structure observed in Fig. 5(b) requires two
resonances. The lower one is well determined by both $\omega \eta$ data
($M = 1981 \pm 23$ MeV) and $\omega \pi \pi$ data ($1973 \pm 24$ MeV).
The upper one is determined better by $\omega \eta$ data.
By comparison with $I = 1, C = -1$ of Ref. 3, it is expected
around 2235 MeV.
In present data it appears somewhat lower, at $2195 \pm 30$ MeV,
though within the combined errors.

For $J^{PC} = 3^{--}$, three states are definitely required.
The lowest contributes a huge $f_2\omega $ signal in
$\pi ^0 \pi ^0 \omega$ data, see Fig. 5(d); it is also quite large in
$\omega \eta$.
This state is expected to lie close to the very well defined $I = 1$
state $\rho _3(1982)$.
That resonance is
defined by extensive $\bar pp \to \pi ^-\pi ^+$ data at
100 MeV/c steps of momentum down to 360 MeV/c (a mass of 1910 MeV) [11];
its mass is $1982 \pm 14$ MeV and the width $188 \pm 24$ MeV.
In present data, the $I = 0$ state optimises at $1944 \pm 16$ MeV
in $\omega \pi ^0 \pi ^0$ and at $1951 \pm 21$ MeV in
$\omega \eta$.
Table 2 quotes an average value $1945 \pm 20$ MeV; the error
allows for the small discrepancy between channels.
It is remarkably narrow in both data sets: $\Gamma = 115 \pm 22$ MeV.
This narrow width is required to fit  a very rapid variation in
differential cross sections between beam momenta of 600 and 900
MeV/c.
However, we warn that this state lies right at the bottom of the
available mass range, so there could be some systematic error in
determining its mass and width.

A $^3G_3$ state is required strongly by both $\omega \eta$ and
$\omega \pi ^0 \pi ^0$ data. It optimises at $M = 2255 \pm 15$
MeV with $\Gamma = 175 \pm 30$ MeV.
It is conspicuous because of its different helicity components from
$^3D_3$, hence different tensor polarisations.
A $^3D_3$ state is also required at a similar mass.
However, this is the least well determined of all the
resonances.
It makes a fairly small contribution to $\omega \eta$ data.
For $\omega \pi ^0 \pi ^0$, a strong contribution is required.
However, the mass drifts up to  2310 MeV with only a small
improvement in log likelihood;
in the earlier analysis of Ref. 2, it optimised instead at
$2180 \pm 40$ MeV.
These instabilities in mass are correlated with flexibility in
fitting the $\omega \sigma$ component in $\omega \pi \pi$.
For the present solution, our final best estimate for the mass (mostly from
$\omega \eta$ data) is $2285 \pm 60$ MeV. For both $\omega \eta$ and
$\omega \pi ^0 \pi ^0$, this $^3D_3$ state is obscured by the very
large low mass contributions from $\omega _3(1945)$.

A small but well determined $^3G_4$ peak is now required in $\omega
\eta$; a $^3G_5$ state is required strongly by both $\omega \eta$ and
$\omega \pi ^0 \pi ^0$ data.
For all $G$-states, the intensity peaks strongly at $\sim 2300$ MeV,
see Figs. 4 and 5.
However, there is a very strong centrifugal barrier in the $\bar pp$
channel.
In consequence, we find anomalously low resonance masses for all
three $G$-states, compared with neighbouring $F$-states, which cluster
from 2260 to 2305 MeV.
We use standard Blatt-Weisskopf centrifugal barrier factors,
which are derived assuming a square well potential.
This may not be a good approximation for very high
partial waves such as $G$-states.

\begin{figure}
\vskip -17mm
\centerline{\epsfig{file=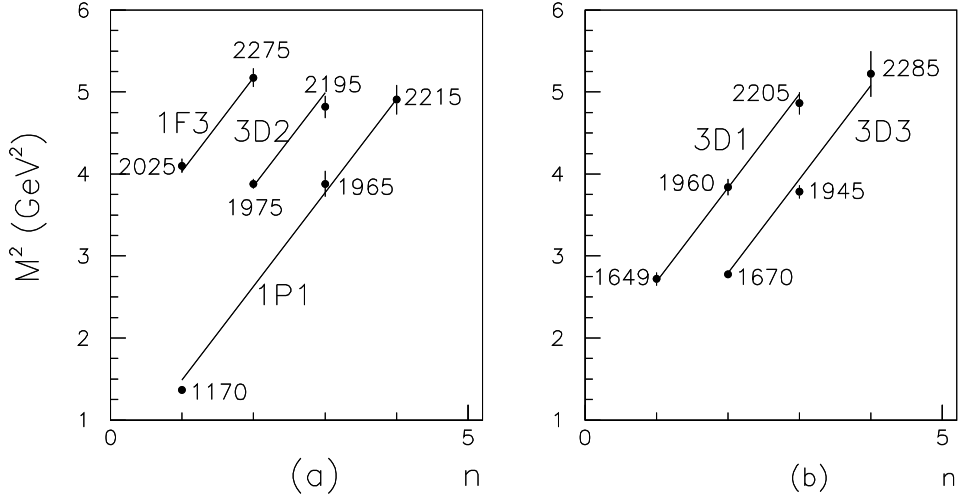,width=15cm}}
\vskip -150.3mm
\centerline{\epsfig{file=CMI0.PS,width=15cm}}
\vskip -71mm
\caption{A comparison of $M^2$ for resonances with straight-line
trajectories against radial excitation number $n$; the slope of 1.143
GeV$^2$ is taken from Ref. [5]. In (b), the $^3D_3$ trajectory is
displaced one place to the right in $n$ in order to resolve it
from $^3D_1$.}
\end{figure}

Fig. 6 shows a plot of observed resonances versus $M^2$, where $M$ is
mass.
They conform closely to straight-line trajectories resembling
Regge trajectories, to which they are related. They are compared in the
figure with a slope of 1.143 GeV$^2$. That slope is determined from
very precise results for $I = 0$ $C = +1$ [5]; for those quantum
numbers, data are available from seven channels including valuable $\pi
^-\pi ^+$ data from a polarised target. Within the errors of present
determinations, there is good agreement with this slope.

\begin{figure}
\centerline{\epsfig{file=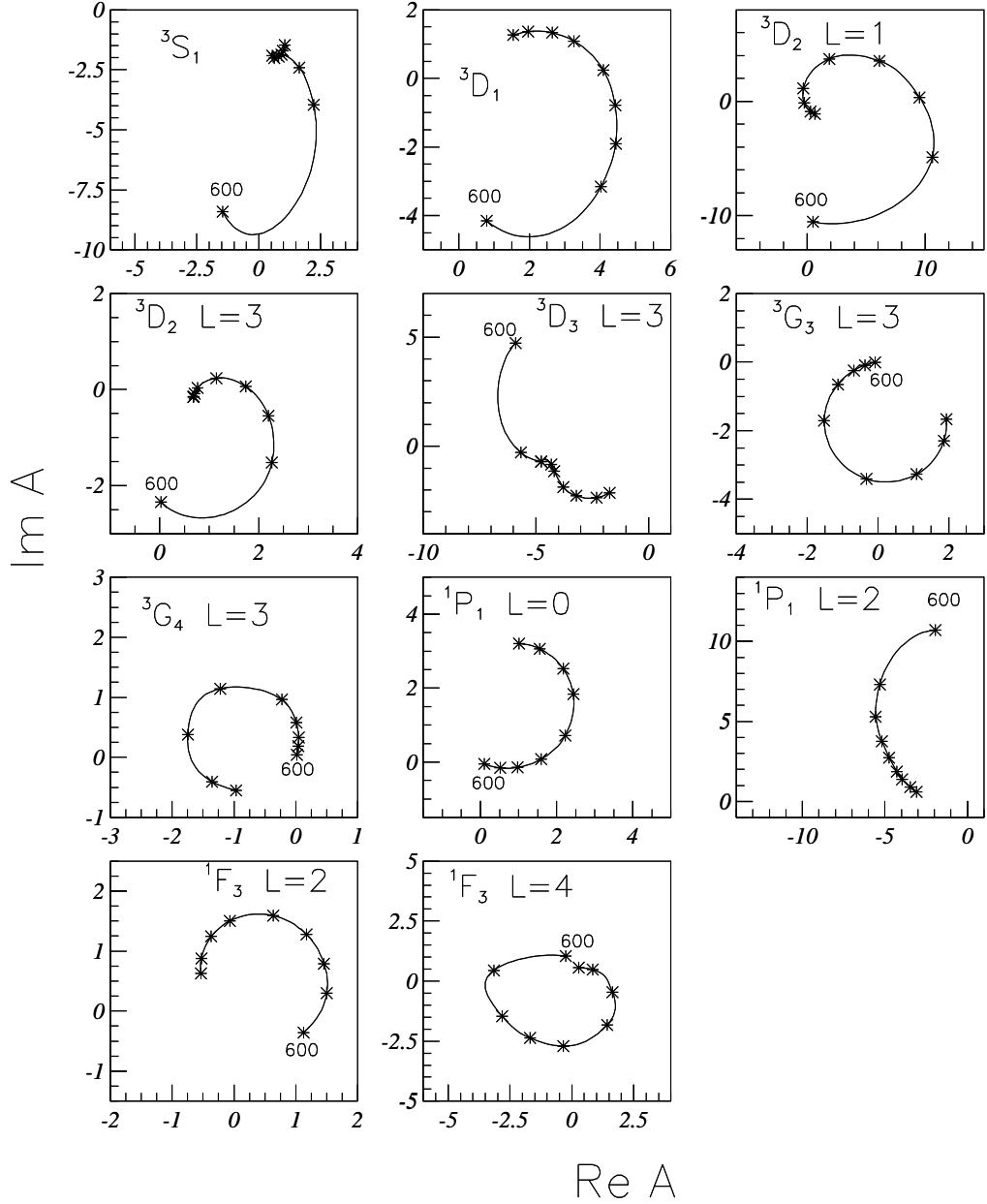,width=15cm}}
\vskip -180.3mm
\centerline{\epsfig{file=F7_WET.PS,width=15cm}}
\caption{Argand diagrams for fitted $\omega \eta $ partial waves.
Crosses show beam momenta 600, 900, 1050, 1200, 1350, 1525, 1642, 1800
and 1940 MeV/c; all move anti-clockwise with increasing beam momentum.}
\end{figure}

Fig. 7 shows Argand diagrams for all partial waves.
A remarkable feature of the vector polarisations of Fig. 2 for
$\omega \eta$ is that they are everywhere small.
The same feature has been observed for
$\omega \pi ^0$.
Vector polarisation depends on the imaginary parts of interferences
between partial waves.
The small observed polarisation requires coherence amongst
all partial waves.
It is discussed in detail in the accompanying paper.

In summary, the new data provide a considerable improvement in
parameters of several resonances.
This arises from more accurate polarisation information
for the $\omega$, leading to better phase determinations.
A new but expected $3^{+-}$ resonance is observed at $2025 \pm 20$
MeV with $\Gamma = 145 \pm 30$ MeV.

\section{Acknowledgement}
We thank the Crystal Barrel Collaboration for
allowing use of the data.
We acknowledge financial support from the British Particle Physics and
Astronomy Research Council (PPARC).
We wish to thank Prof. V. V. Anisovich for helpful discussions.
The St. Petersburg group wishes to acknowledge financial support
from grants RFBR 01-02-17861 and 00-15-96610 and from PPARC;
it also wishes to ackowledge support under the Integration of the
Russian Academy of Science.


\begin {thebibliography}{99}
\bibitem {1} A. Anisovich et al., Phys. Lett. B507 (2001) 23.
\bibitem {2} A. Anisovich et al., Phys. Lett. B476 (2000) 15.
\bibitem {3} A. Anisovich et al., {\it Combined analysis of meson
channels with $I=1$, $C = -1$ from 1940 to 2410 MeV},
accompanying paper.
\bibitem {4} K. Peters, Nucl. Phys. A692 (2001)
295c. \bibitem {5} A. Anisovich et al., Phys. Lett. B 491 (2000) 47.
\bibitem {6} A. Anisovich et al., Phys. Lett. B 517 (2001) 261.
\bibitem {7} A. Anisovich et al., Phys. Lett. B 517 (2001) 273.
\bibitem {8} D.V. Bugg, A.V. Sarantsev and B.S. Zou, Nucl. Phys. B 471,
 (1996) 59.
\bibitem {9} J.E. Augustin et al., Nucl. Phys. B320 (1989) 1.
\bibitem {10} S. Godfrey and N. Isgur, Phys. Rev. D32 (1985) 189.
\bibitem {11} A. Hasan et al., Nucl. Phys. B378 (1992) 3.
\end {thebibliography}
\end {document}